\documentclass[aps,prd,10pt,nofootinbib,twocolumn,eqsecnum,showpacs,showkeys,superscriptaddress,preprintnumbers,altaffilletter,floatfix]{revtex4-1}
\usepackage{ulem}
\usepackage{amssymb}
\usepackage{graphicx}
\usepackage{amsmath}
\usepackage{multirow}
\usepackage{hyperref}
\usepackage{soul}
\usepackage{graphicx}
\usepackage{dcolumn}
\usepackage{amssymb}
\usepackage{amsfonts}
\usepackage{amsbsy}
\usepackage{color}
\usepackage{rotating}
\usepackage[english]{babel}
\usepackage{multirow}
\usepackage{float}

\usepackage{animate}

\usepackage{hyperref}
\hypersetup{
    colorlinks=true,
    linkcolor=blue,
    filecolor=magenta,
    urlcolor=cyan,
    citecolor=cyan
}

\newcommand{\appropto}{\mathrel{\vcenter{
  \offinterlineskip\halign{\hfil$##$\cr
    \propto\cr\noalign{\kern2pt}\sim\cr\noalign{\kern-2pt}}}}}
\begin{document}

\title{Ripped \texorpdfstring{$\Lambda$}~CDM: an observational contender to the consensus cosmological model
}

\author{R. Lazkoz}  
\email{ruth.lazkoz@ehu.eus}
\affiliation{Fisika Saila, Zientzia eta Teknologia Fakultatea, Euskal Herriko Unibertsitatea UPV/EHU, 644 Posta Kutxatila, 48080 Bilbao, Spain}
\affiliation{
EHU Quamtum Center, University of the Basque Country UPV/EHU, PO Box 644, 48080  Bilbao, Spain}
\author{V. Salzano}
\email{vincenzo.salzano@usz.edu.pl}
\affiliation{Institute of Physics, University of Szczecin, Wielkopolska 15, 70-451 Szczecin, Poland}
\author{L. Fern\'andez-Jambrina}
\affiliation{Matemática e Informática Aplicadas, E.T.S.I. Navales, Universidad Politécnica de Madrid, Avenida de la Memoria 4,
E-28040 Madrid, Spain}
\email{leonardo.fernandez@upm.es}
\author{M. Bouhmadi-L\'opez}
\email{mariam.bouhmadi@ehu.eus}    
\affiliation{IKERBASQUE, Basque Foundation for Science, Plaza Euskadi 5
48009 Bilbao, Spain}
\affiliation{Fisika Saila, Zientzia eta Teknologia Fakultatea, Euskal Herriko Unibertsitatea UPV/EHU, 644 Posta Kutxatila, 48080 Bilbao, Spain}
\affiliation{
EHU Quamtum Center, University of the Basque Country UPV/EHU, PO Box 644, 48080  Bilbao, Spain}

\date{\today} 

\begin{abstract}
Current observations do not rule out the possibility that the Universe might end up in an abrupt event, but that fatality might be avoided, as it happens in the  pseudorip dark energy scenario in the form of asymptotically de Sitter behaviours. Different such evolutions may be explored through suitable parameterizations of the dark energy and then confronted to cosmological background data. Here we parameterize a pseudorip cosmological model assuming a particular sigmoid function, and carry an in-depth multifaceted examination of its evolutionary features and statistical performance.  
Further insight into the model's physics is obtained through a phantom scalar field reinterpretation for which the potential is reconstructed in the small and large scale factor regimes.
This depiction of a non-violent final fate of our cosmos seems to be arguably statistically favoured over the consensus $\Lambda$CDM model according to some Bayesian discriminators. These conclusions are drawn using a combination of state-of-the-art low and high redshift cosmological probes.
\end{abstract}


\maketitle

\section{Introduction}
Contradicting though it may seem, our certainty about the current accelerating status of the Universe does not clash with our unpredictability about its final destiny. The data we render as exquisite (as compared with the situation in preceding decades) are in fact not too predictive about the final stages (if this makes sense) of our Universe. Given the wideness of the uncertainty window our data lead to, we cannot ascertain now whether the Universe will expand peacefully for ever or whether a doomsday will occur. More precisely, the evolution of the Universe may be awkward enough to show an observational preference for phantom dark energy at present which might healthily evolve to finally avoid the characteristic Big Rip singularity.

Solving this riddle implies determining with enough accuracy the evolutionary features of dark energy \cite{Huterer:2017buf}. Reliable constraints are typically obtained for the present value  of the  equation of state parameter of the dark energy. Unfortunately, the derivative of that parameter remains refractary to accuracy, and the associated observational errors  remain too large to offer satisfactory conclusions. This
lets us, thus, be quite certain about the current accelerated expansion of the universe, but does not allow to say whether acceleration will increase or decrease. Note that this vagueness is  common to all phenomenological parametrizations the community has come up with so far when they are subject to observational tests.  Needless to say, models proposed from the modified gravity perspective do not remedy this precarious issue.

But we can turn the tables and see this as an opportunity to learn more about how borderline dark energy models can display dramatically amplified different evolutionary features. Indeed, just a tiny percentual change in the current value of the equation state parameter around the cosmological constant frontier may lead to a Universe with a violent fate.  In the light of this,  a lot of effort is being put into using astrophysical and cosmological probes to narrow down  the parameters which enter the specific functional form of the $p/\rho$ ratio for each  dark energy equation proposal.

These riddles provide a good motivation to establish a dialogue between two routes that have mainly remained unconnected.  The first one embraces taxonomical studies of the cosmological features of dark energy models with  a phantom nature and in a broad sense, and in particular those of the scenarios evolving towards a de Sitter asymptotic future.   The second is the much broader research area of dark energy parametrizations in general from the observation point of view to address  either models emerging from a purely phenomenological perspective or with some fundamental motivation.

The  inescapable uncertainty window offered by cosmological constraints on the dark energy window leave room to accommodate new parametrizations encompassing the most intriguing features of those two worlds. In fact, specifically we propose a a pseudorip scenario and perform a thorough multifaced test on it. According to the definition, and despite what their name may seem to indicate, pseudorip cosmologies belong to the phantom realm for their whole history, just to reach a final de Sitter stage which prevents the fatal consequences on compact objects which phantom cosmologies are responsible for.

The scheme on the paper is as follows. Section \ref{sec:context} discusses the state of the art for the two areas that provide context for our model: cosmological features on the one hand and dark energy parametrizations on the other. Section \ref{sec:description} describes the model and reveals associated basic background evolutionary features such as the Hubble parameter, the equation of state parameter, and scale factor asymptotic behavior. Morever, we discuss a possible connection to fundamental physics and debate about its reach and applicability. In section \ref{sec:Data} we provide a detailed description of the cosmological datasets considered and the statistical framework needed to draw conclusions when connecting them to our proposal. In Section \ref{sec:results} we present our results with a pertinent discussion. Finally, in Section \ref{sec:conclucions} we draw our main conclusions emphasizing how intriguing and promising possibilities are opened by this pseudorip scenario and hypothetical future ones.

\section{Context for the model}
\label{sec:context}

\subsection{Cosmological futures}

One of the typical features of  phantom dark energy ($p/\rho\lesssim-1$)  is that it does not respect the null energy condition ($\rho+p\le 0$) \cite{Carroll:2003st}.  As this type of content of the universe is often observationally allowed (and sometimes even favoured \cite{Escamilla:2023oce, Vagnozzi:2019ezj}), the violation of the null energy condition enforced a fresh open-mindedness  about the dark energy nature. This paved the way to the discovery of unusual fates for the universe. The profuse work that followed in this area was instigated by two main references concerned with the behaviour of the scale factor, the energy density and the pressure. Note that using the standard equation of energy conservation it can be shown that phantom dark energy is a blueshifting component.

Phantom dark energy itself
was the first representative of the now populated taxonomy of abrupt cosmological endings.
In the influential work \cite{Caldwell:2003vq} the unavoidable Big Rip singularity of phantom universes was introduced,  and the consequent finite time blow up of the scale factor, the energy density and the pressure were discussed. Such singularities exclude by definition the
$p=-\rho$ possibility. In the case that the rip arises when that condition is reached at the final stage of the evolution of the model, another type of singularity appears, which has been dubbed Grand Rip \cite{Fernandez:2014} or Type -1 \cite{Fernandez:2016}.

In the second inspiring reference Barrow presented a different type of cosmological milestone in which  the unbounded instantaneous growth would affect the pressure, whereas the energy density and the scale factor would not be pathological \cite{Barrow:2004xh}.
The singularity is reached at a finite time in this case too. These were originally referred to as sudden singularities, but this denomination has gotten somewhat blurred by it being used perhaps a bit too loosely.

The Big Rip and Barrow singularities correspond respectively to types I and II in the classification of \cite{Nojiri:2005sx}, which is based on the behaviour of the Hubble factor and its  derivatives. This is certainly a convenient first approach for a large-scale evolution examination of these models. This classification has been progressively enlarged to the final list \cite{Fernandez:2023}.

Fuelled by a conspicuous imagination, theoretical cosmologists have coined names for a number of particular cases of type II singularities. Specifically, Big Boost singularities are those with a positive  second derivative of the scale factor  \cite{Barvinsky:2008rd}, whereas their negative counterparts are of no concern to us as we are concentrating on universes which accelerate as long as they remain to exist once they start doing so.

Another worth mentioning type of singularity \cite{Bouhmadi-Lopez:2007xco} is  the type III one in the sorting mentioned above. This corresponds to both the energy density and pressure displaying an unbounded growth as a finite value of the scale factor is approached and all this happening at a finite time, so it is yet another flavour of what we call a rip in a broad sense. 

A related singularity in the sense it can be regarded as the dual of the former is type IV \cite{Nojiri:2005sx}. The
pressure and energy density  evolve to null values in finite time and for a bounded value of the scale factor, whereas derivatives of $H$ higher than order one will diverge. 
These models are not really of interest for us as this type of dark energy ceases to produce acceleration at some point. In the case  the derivatives of $H$ do not diverge, only  the barotropic index $w$ does (recall the standard definition $w=p/\rho$). These singularities have been referred to as Type V or barotropic index singularities (in a more behaviour revealing denomination) \cite{Dabrowski:2009,Fernandez:2010}.


Finally, we might consider directional singularities \cite{Fernandez:2007}, which have been added to the list as Type $\infty$ singularities \cite{Fernandez:2014}.  Such singularities appear at infinite coordinate time, but finite proper time along causal geodesics, and are not experienced by all observers. Due to their pathological features, we do not consider them here either.

Along this route there is another scheme to put together different scenarios under the umbrella of the ``rip" denomination. In such models the inertial force on certain mass as measured by an observer separated a given  comoving distance reaches either an infinite or a very large value. As that force is proportional 
to the  combination $ H+\dot H$ its behavior offers an immediate test to examine and sort cosmological models.

``Rip" scenarios may be singular or not, though.  For the little rip \cite{Frampton:2011,Frampton:2012a} and the little sibling of the big rip \cite{Bouhmadi-Lopez:2014cca} scenarios, the energy density and pressure grow unboundedly  in the future and their evolution lasts for an infinite amount of time. This makes them non-singular (unlike the traditional Big Rip or the recently discovered Grand Rip). The little sibling is milder in the sense that the cosmic time derivative of the Hubble ratio does not diverge.

Another non-singular case is the pseudorip scenario \cite{Frampton:2012} in which the energy density and the pressure tend to a finite value and the destruction of structures occurs for binding forces below some particular threshold. But as their dark energy is phantom-like all the way to just the end the inertial force can be expected to be huge. 

Pseudorip models belong to  the type  will precisely focus on. The motivation is that in what regards their functional form they  represent (as we shall see) a smooth departure from the $\Lambda$CDM case while they are at the same time able to depict a long-standing preceding phantom epoch. We present here a new parametrization reproducing such an evolution and perform a state-of-the-art observational analysis following some interesting theoretical insights.\\

\subsection{Dark energy parametrizations reviewed}
In very broad terms, dark energy parametrizations aim at 
smoothing out an array of observational data, which are basically unstructured and scattered. Mathematically speaking this procedure is well-defined and expected to be reasonably informative.  Besides,  in practice, that route is technically not as challenging as non-parametric reconstructions. 

Just for review purposes we can recall that three have been the main ways to reconstruct the dark energy with non-parametric procedures: Gaussian processes \cite{Seikel:2012uu,Holsclaw:2010nb,Holsclaw:2010sk,Holsclaw:2011wi,Shafieloo:2012ht,
Liao:2019qoc,Keeley:2020aym,Hwang:2022hla}, principal components  \cite{Crittenden:2005wj,
Huterer:2002hy, Ruiz_2012}
and  local regressions inside sliding windows 
\cite{Daly:2003iy, Daly:2007dn, Daly:2006ax, Daly:2004gf, Abdalla:2022yfr,Montiel:2014fpa, Rana:2015feb}

However, we are concerned here by parametrizations of  
dark energy fuelled cosmological backgrounds, and the literature offers examples galore. Note that our focus is on cases in which the dark energy is conserved and therefore only interacts gravitationally with other components (dark matter, radiation, etc.) as it happens in alternative dark energy models (see \cite{Lima:2004cq,arun2017dark}).
We can distinguish three main families depending on the quantity to being ultimately fitted: the comoving distance (or relatedly the luminosity distance), the Hubble factor, and the dark energy equation of state parameter $w$  (quite possibly the specific notation  $p=w\rho$ was first ever used in \cite{Turner:1997npq}).  

 For the luminosity functions study cases come in many flavours. Early works resorted to  rather simple parametrization ans\"atze
 \cite{Saini:1999ba} whereas more modern work rest upon different types of truncated polynomial  expansions  
\cite{Capozziello:2011tj, Visser:2004bf,Cattoen:2008th, Guimaraes:2010mw,Cai:2011px,Aviles:2012ay}, of which Padé approximants are but one of the scenarios explored
 \cite{Zaninetti:2016fju,Gruber:2013wua}. Less known more recent works resort to the demanding holomy perturbation methods (HPM) on their own \cite{Shchigolev:2015snv} for combined with Padé approximants \cite{Yu:2021lzc}.   
 Generally speaking, one of the problems of this type of parametrizations based on the luminosity distance learning about the kinematics   two or differentiation steps will be necessary if one wishes to go beyong $w$ and get cosmographic insights. The resulting functions will  typically be plagued with correlations among the observationally parameters and a very likely amplification of the uncertainties. In this framework the method of orthogonalized logarithmic polynomials has emerged recently \cite{Bargiacchi:2021fow} and it seems to 
 be able to remove covariance between luminosity distance parameters. However additional insight will be needed to see how this translates into other magnitudes characterizing dark energy evolution in the context of the model. 

 The opposite situation is the one that kicks-off from specifying $w$. A double integral is necessary to obtain the associated form of the luminosity distance in order to use type Ia supernovae data for the tests. Integrating twice may iron out ups and downs in the underlying $w$ and the final fit may be unable to capture them. For this reason considering more than two parameters characterizing $w(z)$ is frowned upon. Besides, errors in the second parameter (say, $w'(z)\vert{z=0}$) are sensitive to the specific form of $w(z)$  so conclusions will be compromised if too much focus is set on a specific model \cite{Colgain:2021pmf}. A comprehensive list of authors which have produced interesting such parametrizations beyond the ubiquitous CPL model is
\cite{Wetterich:2004pv,Jassal:2004ej,Feng:2012gf,Nesseris:2004wj,Linder:2006sv,Cooray:1999da,Ma:2011nc,Mukherjee:2016trt, Gong:2005de, Lee:2005id, Upadhye:2004hh,Gerke:2002sx, Lazkoz:2010gz,Sendra:2011pt,feng2011new,Efstathiou:1999tm,Barboza:2011gd,Yang:2021eud}.

 Is there then an alternative to pursue? Indeed, we believe fits of the dark energy density itself offer a good compromise in this sense, as we have to go just one level up to use luminosity data, and one level down to extract basic kinematic conclusions. 

The evident profusion of available dark energy prescriptions might make doubts arise about the interest in discussing a new one in detail. However, dark energy parametrizations with ``peculiar" fates other than Big Rip singularities are still somehow uncharted territory. Therefore  there seems to be room for new proposals. Even more so when the intention is to address them  carrying out an unprecedentedly thorough observational analysis. 

\section{Model description}
\label{sec:description}

\subsection{A new pseudorip dark energy scenario}

The phenomenological addition of a non-redshifting term in the Einstein equations is almost as old as General Relativity itself. This term remained a subject of theoretical and empirical interest even in the epochs in which its contemplation was frowned upon.

Interestingly a term like this emerging from a purely mathematical argument was introduced to try and reproduce the then accepted  physical description of the universe, even though there was no fundamental physics to support the presence of such term. Yet it contributed enormously to understand the kind of kinematics such a term can induce and eventually gave rise to a description of the universe which has been shown to be quite proficient.

We are now in a similar situation, where no deep understanding of the cause of the observed acceleration exists, yet we need to make the most of exquisite observational evidence.

In fact progress in solid research subareas in the field  is being made towards  answering  fundamental open questions about the evolution of the universe even in the lack of a strong connection with fundamental physics.  Many advances in our understanding of the evolution of the universe are connected to the many different parametrizations that have been proposed so far often without underlying theoretical incentives. Just to mention one of those questions an apparently naïve one is what is the age of the universe. As this is known to be critically dependent on the specific dark energy model, having broad catalogues will certainly help us make progress by identifying patterns and robust results. On top of that, phenomenological parametrizations have historically offered help to design future cosmological surveys through forecast informing us of the constraining power of those missions. Thus, proposing new parametrizations to explore not so chartered aspects of dark energy evolution such as the final destiny of the universe is still a commendable pursuit. 

With that extra motivation sheet we notice that the pseudorip behaviour we are after can be  modeled with  sigmoid functions (which in some contexts are dubbed as Ridge (activation) functions \cite{pinkus_2015}). Remember that in a pseudorip scenario phantom dark energy is tamed, so to speak, and does not display its typical unbounded growth, but it rather interpolates between an early phantom quintessence regime and a late de Sitter one. 

Precisely we propose a phenomenological dark energy density in which the energy density and the scale factor are related through the Gudermannian function 
\begin{equation}
gd(x)=\arctan(\sinh(x))=2\arctan(\tanh(x/2))
\end{equation}
so that for this blueshifting component we write
\begin{equation}\label{eq:density_lambda}
\rho=\frac{2}{\pi}\rho_
{\Lambda}\,{\rm gd}(\lambda a^{\eta}),
\end{equation}
with positive constants $\rho_{\Lambda},\lambda,\eta.$
 As customary, we will make the choice
$a_0=1$ for the current value of the scale factor, and this makes $\lambda$ dimensionless.

The Gudermannian function has been used in field theory to obtain solutions in the classical sine-Gordon theoretical framework \cite{Mondaini:2014sya}. From the operational perspective the ability of this sigmoid function to relate hyperbolic functions to trigonometric functions without resorting to complex numbers keeps expressions compact despite the
backs and forths to present quantities in terms of $a$ or $\rho$. This makes it an interesting choice over other sigmoid functions, especially because this is a first attempt to carry out a thorough observational analysis of a dark-enery parametrization with a pseudorip nature.

The scale factor becomes infinite when $t\to\infty$. This is a pseudo-rip cosmological milestone, also characterized for a finite  $H$ and  null $\dot H$ at infinite $a$, which is given in terms of $\rho$ by
\begin{eqnarray}
a=\left(\lambda^{-1}gd^{-1}\left(\frac{\pi  \rho}{2
  \rho_{\Lambda}}\right)\right)^{\frac{1}{\eta }}.\\\nonumber
\end{eqnarray}
A better grasp of how this scale factor behaves near the event is given by
\begin{equation}
    \lim_{\rho \to  \rho_{\Lambda}^{-}}\frac{a}{(-\log\vert\rho_{\Lambda}-\rho  \vert)^{1/\eta}}=\lambda^{-1/\eta} . 
\end{equation}

It is easy to check that there is a pseudorip for large $a$, since there
\begin{equation}\frac{\rho(a)}{\rho_{\Lambda}}\approx 1-\frac{4}{\pi}e^{-\lambda 
a^{\eta}}. 
\end{equation}
In order to obtain $a(t)$ in an approximate way, we need to integrate
\begin{equation}\sqrt{3}\frac{\dot a}{a}=\sqrt{\rho}\approx 
\sqrt{\rho_{\Lambda}}\left(1-\frac{2}{\pi}e^{-\lambda 
a^{\eta}}\right),\end{equation}
where we have set $8\pi G=1$. The result is
\begin{equation}\frac{\sqrt{\rho_{\Lambda}}}{\sqrt{3}}t\approx 
\ln a -\frac{2}{\pi\eta}\mathrm{Ei}\,(\lambda a^{\eta}),\end{equation}
where the exponential integral $\mathrm{Ei}$ is defined as
\begin{equation}\mathrm{Ei}\,(x):=\int_{x}^{\infty}\frac{e^{-t}}{t}dt\approx 
e^{-x}\ln(1+1/x)\end{equation} for large $x$.

We see hence that for large $a$, the dominant contribution to $t$ is 
that of $\ln a$ and so the milestone is reached for infinite 
coordinate time $t$.

Furthermore,
\begin{equation}\frac{\pi}{2}\frac{\sqrt{3}\dot H(a)}{\sqrt{\rho_{\Lambda}}}\approx
\lambda \eta a^{\eta}e^{-\lambda a^{\eta}}H(a)\end{equation}
tends to zero for large $a$ and so do further derivatives of $H$, as it happens in pseudorip milestones.

Note that $\rho_{\Lambda}$ plays the same role of the critical energy density discussed in \cite{Fernandez:2009}.

The case of negative $\eta$ is simpler, though it is not favoured by observations. In this case, for small $x$, $\mathrm{gd}\,(x)\approx x$ and $\rho(a)$ behaves as $a^\eta$. Hence for large $a$, the energy density, the Hubble constant and its derivatives vanish at an infinite coordinate and proper time. 

For an observational test at the background level and assuming spatial-flatness we would simply need this starting point:
\begin{eqnarray}
H^2&=&H_0^2\left[\Omega_m(1+z)^3+\Omega_r(1+z)^4\right.\nonumber\\&+&\left. \frac{2\,\Omega_{\Lambda}}{\pi} gd\left(\lambda\, (1+z)^{-\eta}\right)\right],
\end{eqnarray}
where $\Omega_{\Lambda} =\rho_{\Lambda}/ (3H_0^2)$ and the parameter $\lambda$ is now defined as
\begin{equation}
\lambda=gd^{-1}\left( \frac{\pi({1-\Omega_m-\Omega_r})}{2\Omega_{\Lambda}} \right)\, ,
\end{equation}
to ensure the proper normalization. In this regard, we can reasonably expect
${(1-\Omega_m-\Omega_r)/\Omega_{\Lambda}}\sim {\cal O}(1).$ This result along with the properties of the  inverse Gudermannian function lets us finally infer $\lambda\sim {\cal O}(1)$.

It is also of interest to see that
 the standard conservation equation
 leads to the equation of state
\begin{equation}
     p=-\rho-\frac{2\eta}{3\pi}{\rho_{\Lambda}}\,\,gd^{-1}\left(\frac{\pi\rho}{2\rho_{\Lambda}}\right)\cos\left(\frac{\pi  \rho}{2\rho_{\Lambda}}\right)\nonumber\end{equation}
     which is valid for $\rho\leq \rho_{\Lambda}.$
     This gives a quiessence phantom behavior at kick-off, as 
     $p=-(1+{\eta}/{3})\rho+{\cal O}(\rho^2),$
whereas the final behaviour is clearly that of a cosmological constant
     as $\lim_{\rho\to\rho_{\Lambda^{-}}}p/\rho=-1.$

     Alternatively we can write
\begin{equation}
 p=-\frac{2}{\pi} {\rho_{\Lambda}}
\left({\frac{\eta\lambda}{3}
   a^{\eta } \text{sech}\left(\lambda 
   a^{\eta }\right)}+{\rm gd}(\lambda a^{\eta})\right).    \end{equation}

Now, regardless of the sign of $\eta$ we see that the equation of state parameter, generally defined as
\begin{equation}
w(a) = -1 - \frac{a\rho'(a)}{3\rho(a)}\,,
\end{equation}
when evaluated at present time has this value:
\begin{equation}
w_0=-1-\frac{\eta  \lambda  \text{sech}(2 \lambda )}{3 \arctan(\sinh (\lambda ))}.
\end{equation}
This would offer (if needed) additional insight into the values of $\eta$ and $\lambda$ leading to 
acceleration at present, that is, $w_0<-1/3$.

\subsection{Connection to fundamental physics}

Reinterpreting our dark energy component in terms of a phantom scalar field in the $\eta>0$ case can be an exercise of interest
and give our proposal further support.  As customary
we set
\begin{eqnarray}
&&\rho=-\frac{\dot \phi^2}{2}+V(\phi),\\
&&p=-\frac{\dot \phi^2}{2}-V(\phi),
\end{eqnarray}
which leads to 
\begin{equation}\dot \phi^2=-(w+1)\rho
.\end{equation}
As derivatives with respect to $a$ render the equations simpler we reformulate the problem using them. Then, taking Eq. into account we get
\begin{equation}
\phi'=
\sqrt{\frac{\rho'}{3}aH^2} %
\end{equation}
This result can be seen to have full compatibility with those in \cite{Battye:2016alw}.

Obviously, practically in any case the high non-linearity of the functions involved will at some point hinder a complete resolution of the problem, either because integrals 
cannot be done analytically or because writing the scale factor as a function of $\phi$ after an inversion is not possible either. Thus, in order to make some progress 
approximations are needed, leading to separate valid results  for small and large scale factor values.\\

\subsection{Potential reconstruction near the de Sitter epoch}
When $a$ is very large and we are close to the de Sitter epoch in the asymptotic future we have
\begin{equation}
    H^2\sim\rho_{\Lambda}/3
\end{equation}
which, according to our previous results, leads to
\begin{equation}\phi'\simeq
\sqrt{\frac{4\eta{\lambda}}{\pi}{ a^{\eta-2}}e^{-\lambda a^{\eta}}}
\end{equation}
This can be integrated to give 
$$\phi\sim\sqrt{\frac{8}{\eta }}\text{erf}\left(\sqrt{\frac{\lambda}{2}}a^{\eta /2}\right)$$
where a shift has been done which gives  us $$\phi_{\infty}= \lim_{a\to\infty}\phi=\sqrt{8/\eta}\sim{\cal O}(1)\,.$$ This is an upper limit for $\phi$ and its existence allows the scalar field to be compatible with the swampland distance conjecture \cite{Ooguri:2018wrx,Storm:2020gtv}.
On the other hand in this regime $p(a)\simeq -\rho(a)$ and therefore, as this corresponds to a stage dominated by the potential energy we can write $V(\phi)\sim \rho $
and subtituting the expression for 
$\phi$ we finally get:
$$V(\phi)=\rho_{\Lambda}\left(1-\frac{4}{\pi }{ e^{-2 \left(\text{erf}^{-1}\left(\sqrt{\displaystyle\frac{\eta}{8}}\phi\right)\right)^2}}\right).$$
Formally the potential has an inverted bell shape with truncated wings (it is a bottom-up Gaussian with a cumbersome argument). Obviously, the function  is symmetric with respect to $\phi$, and its wings are truncated of the already mentioned upper limit of $\phi$, which in turn gives
$V\le \rho_{\Lambda}$. The late-time acceleration we are describing is linked to a slow uphill roll of the phantom scalar field in this concave potential, as customary in models of this sort  (see \cite{PhysRevD.74.083501,Dutta:2009dr} for generic discussions). Note, in any case that this potential is only valid in the specific close-to-de-Sitter regime we are depicting

\subsection{Phantom quintessence-like epoch}
Regardless of the epoch under consideration we have
$
V=\rho+p$,
but given the nearly constant character of the equation of state parameter
 for low scale factor values
 it follows that
 $V\propto\rho$
 gives us a very good approximation.
Now, for $\eta>0$ we can write the following for expression low scale factor values:
$$\rho(a)=\rho_{\lambda}a^{\eta}+{\cal O}(a^{3\eta}).$$
Thus, taking into account that in this regime approximately $H^2\propto a^{-4}$ we can now write (again, at leading order)
\begin{equation}
\phi'\propto a^{1+\eta/2}.
\end{equation}
Naturally, upon integration and a simple substitution we conclude that 
\begin{equation}
V\propto\rho\propto
\phi^{{2\eta}{/(4+\eta)}}.
\end{equation}
Our result is compatible with the known requirement on the positivity of the exponent to match a positivity in the energy density, unlike what happens with standard (non-phantom) scalar fields \cite{SARIDAKIS2009116}.

\section{DATA}
\label{sec:Data}

For our statistical analysis we will consider a number of different types of cosmological probes, which we briefly describe in the next sections. The respective data compilations are up-to-date.

\subsection{Pantheon+ SNeIa}

The most updated Type Ia Supernovae (SNeIa) data collection is the Pantheon+ sample \cite{Scolnic:2021amr,Peterson:2021hel,Carr:2021lcj,Brout:2022vxf} spanning the redshift range $0.001<z<2.26$. 

It compiles 1701 light curves of 1550 spectroscopically confirmed Type Ia supernovae (SNe Ia), and they are a key piece for distance-ladder analyses to infer cosmological parameters. The significant low redshift increase from the first Pantheon analysis (1048 SNe) to include now $z<0.01$ Cepheid-calibrated data offers the unprecedented possibility to constrain both the Hubble constant (H0) and the dark energy equation-of-state parameter with SNe.

The $\chi^2_{\rm SN}$ will be defined as
\begin{equation}\label{eq:chi_sn}
\chi^2_{\rm SN} = \Delta \boldsymbol{\mathcal{\mu}}^{\rm SN} \; \cdot \; \mathbf{C}^{-1}_{\rm SN} \; \cdot \; \Delta  \boldsymbol{\mathcal{\mu}}^{\rm SN} \;,
\end{equation}
where $\Delta\boldsymbol{\mathcal{\mu}} = \mathcal{\mu}_{\rm theo} - \mathcal{\mu}_{\rm obs}$ is the difference between the theoretical and the observed value of the distance modulus for each SNeIa and $\mathbf{C}_{SN}$ is the total (statistical plus systematic) covariance matrix. The theoretical distance modulus is
\begin{equation}\label{mu_theo}
\mu_{\rm theo}(z_{\rm hel},z_{\rm HD},\boldsymbol{p}) = 25 + 5 \log_{10} [ d_{L}(z_{\rm hel}, z_{\rm HD}, \boldsymbol{p}) ]\; ,
\end{equation}
where $d_L$ is the luminosity distance (in Mpc) is
\begin{equation}
d_L(z_{\rm hel}, z_{\rm HD},\boldsymbol{p})=(1+z_{\rm hel})\int_{0}^{z_{\rm HD}}\frac{c\,dz'}{H(z',\boldsymbol{p})} \,,
\end{equation}
and: $H(z)$ is the Hubble parameter (cosmological model dependent); $c$ is the speed of light; $z_{\rm hel}$ is the heliocentric redshift; $z_{\rm HD}$ is the Hubble diagram redshift \cite{Carr:2021lcj}; and $\boldsymbol{p}$ is the vector of cosmological parameters.

The observed distance modulus $\mu_{obs}$ is
\begin{equation}\label{mu_obs}
\mu_{\rm obs} = m_{B} - \mathcal{M}\; ,
\end{equation}
with $m_{B}$ the standardized SNeIa blue apparent magnitude, and $\mathcal{M}$ the fiducial absolute magnitude calibrated by using primary distance anchors such as Cepheids. While in general $H_0$ and $\mathcal{M}$ are degenerate when SNeIa alone are used, the Pantheon+ sample includes $77$ SNeIa located in galactic hosts for which the distance moduli can be measured from primary anchors (Cepheids), which means that such a  degeneracy can be broken and $H_0$ and $\mathcal{M}$ can be constrained separately. Thus, the vector $\Delta\boldsymbol{\mathcal{\mu}}$ will be
\begin{equation}
\Delta\boldsymbol{\mathcal{\mu}} = \left\{
  \begin{array}{ll}
    m_{\rm B,i} - \mathcal{M} - \mu_{\rm Ceph,i} & \hbox{$i \in$ Cepheid hosts} \\
    m_{\rm B,i} - \mathcal{M} - \mu_{\rm theo,i} & \hbox{otherwise,}
  \end{array}
\right.
\end{equation}
with $\mu_{\rm Ceph}$ being the Cepheid calibrated host-galaxy distance
provided by the Pantheon+ team.

\subsection{Cosmic Chronometers}

Early-type galaxies which both undergo passive evolution and exhibit a characteristic feature in their spectra, i.e. the $4000$ {\AA}  break are generally defined as cosmic chronometers (CC). Stemming from the original idea in \cite{Jimenez:2001gg} the use of that type of galaxies and (more recently) that specific characteristic has lead to their extensive use as ``clocks'' \cite{Moresco:2010wh,Moresco:2018xdr,Moresco:2020fbm,Moresco:2022phi} and can provide measurements of the Hubble parameter $H(z)$ without specific (or within very general) cosmological assumptions \cite{Moresco:2012by,Moresco:2012jh,Moresco:2015cya,Moresco:2016nqq,Moresco:2017hwt,Jimenez:2019onw,Jiao:2022aep}. While not directly constraining the Hubble constant, the information whic CC encode about the cosmological background evolution makes them an optimal testbench for dark energy models.

The most updated sample of CC is from \cite{Jiao:2022aep} and covers the redshift range $0<z<1.965$. The corresponding $\chi^2_{\rm H}$ can be written as
\begin{equation}\label{eq:chi_cc}
\chi^2_{\rm H} = \Delta \boldsymbol{\mathcal{H}} \; \cdot \; \mathbf{C}^{-1}_{\rm H} \; \cdot \; \Delta  \boldsymbol{\mathcal{H}} \;,
\end{equation}
where $\Delta \boldsymbol{\mathcal{H}} = H_{\rm theo} - H_{\rm data}$ is the difference between the theoretical and observed Hubble parameter, and $\mathbf{C}_{\rm H}$ is the total (statistical plus systematics) covariance matrix \cite{Moresco:2020fbm}.

\subsection{Gamma Ray Bursts}

Gamma-ray bursts (GRBs) are high-powered  astrophysical objects which are now believed to offer a good complementarity to type Ia supernovae. The gamma ray bursts (GRBs) ``Mayflower'' sample \cite{Liu:2014vda}, overcomes the well-known circular issue of calibration and ``standardization'' of GRBs by relying on a robust cosmological model independent procedure (by using Padé approximation). It is made of 79 GRBs in the redshift interval $1.44<z<8.1$ for which the authors provide the distance moduli. The $\chi_{\rm G}^2$ is defined exactly like in the SNeIa case, Eq.~(\ref{eq:chi_sn}), but in this case we cannot disentangle between the Hubble constant and the absolute magnitude, so that we have to marginalize over them. Following \cite{Conley:2011ku} it becomes
\begin{equation}\label{eq:chi_grb}
\chi^2_{\rm GRB}=a+\log d/(2\pi)-b^2/d\,,
\end{equation}
with $a\equiv \left(\Delta\boldsymbol{\mathcal{\mu}}_{\rm G}\right)^T \, \cdot \, \mathbf{C}^{-1}_{\rm G} \, \cdot \, \Delta  \boldsymbol{\mathcal{\mu}}_{\rm G}$, $b\equiv\left(\Delta \boldsymbol{\mathcal{\mu}}_{\rm G}\right)^T \, \cdot \, \mathbf{C}^{-1}_{\rm G} \, \cdot \, \boldsymbol{1}$ and $d\equiv\boldsymbol{1}\, \cdot \, \mathbf{C}^{-1}_{\rm G} \, \cdot \, \boldsymbol{1}$, being $\mathbf{C}_{\rm G}$ the covariance matrix and $\boldsymbol{1}$ the identity matrix. 

\subsection{Cosmic Microwave Background}

The Cosmic Microwave Background (CMB) analysis is not performed using the full power spectra provided by the latest release of the \textit{Planck} satellite \cite{Planck:2018vyg}, but rather using the CMB shift parameters defined in \cite{Wang:2007mza} and derived from the latest \textit{Planck} $2018$ data release in \cite{Zhai:2019nad}.  
These quantities are also referred to in the literature as CMB distance priors. According to the comprehensive discussion  in \cite{Zhai:2019nad}, the CMB shift parameters represent one of the least dependent-on-model datasets that can be derived from the CMB power spectrum. In fact they embody almost all the CMB related information which is relevant for investigating dark energy. Additionally, those priors can contribute to a more comprehensive understanding of the CMB restrictions on model parameters, encompassing uncertainties and correlations with other cosmological measurements, without the need to calculate the complete linear perturbation theory CMB quantities. Interestingly, in \cite{Planck:2018vyg} it was also shown that significant differences between the two approaches become manifest when a non-spatially flat geometry is considered, which is not our case. Even more importantly, the use of CMB distance priors was granted approval for cases not stemming from a modified gravity theory that induces consistent alterations to the growth of perturbations by the {\it Planck} collaboration itself \cite{Planck:2015bue}.

The $\chi^2_{\rm CMB}$ is defined as
\begin{equation}
\chi^2_{\rm CMB} = \Delta \boldsymbol{\mathcal{F}}^{\rm CMB} \; \cdot \; \mathbf{C}^{-1}_{\rm CMB} \; \cdot \; \Delta  \boldsymbol{\mathcal{F}}^{\rm CMB} \; ,
\end{equation}
where the vector $\mathcal{F}^{\rm CMB}$ corresponds to the quantities:
\begin{eqnarray} \label{eq:CMB_shift}
R(\boldsymbol{p}) &\equiv& \sqrt{\Omega_m H^2_{0}} \frac{r(z_{\ast},\boldsymbol{p})}{c}, \nonumber \\
l_{a}(\boldsymbol{p}) &\equiv& \pi \frac{r(z_{\ast},\boldsymbol{p})}{r_{s}(z_{\ast},\boldsymbol{p})}\,,
\end{eqnarray}
in addition to a constraint on the baryonic content, $\Omega_b\,h^2$, and on the dark matter content, $(\Omega_m-\Omega_b)h^2$. In Eq.~(\ref{eq:CMB_shift}), $r_{s}(z_{\ast})$ is the comoving sound horizon evaluated at the photon-decoupling redshift, i.e.
\begin{equation}\label{eq:soundhor}
r_{s}(z,\boldsymbol{p}) = \int^{\infty}_{z} \frac{c_{s}(z')}{H(z',\boldsymbol{p})} \mathrm{d}z'\, ,
\end{equation}
with the sound speed given by
\begin{equation}\label{eq:soundspeed}
c_{s}(z) = \frac{c}{\sqrt{3(1+\overline{R}_{b}\, (1+z)^{-1})}} \; ,
\end{equation}
the baryon-to-photon density ratio parameters defined as $\overline{R}_{b}= 31500 \Omega_{b} \, h^{2} \left( T_{\rm CMB}/ 2.7 \right)^{-4}$ and $T_{\rm CMB} = 2.726$ K. The photon-decouping redshift is evaluated using the fitting formula from \cite{Hu:1995en},
\begin{eqnarray}{\label{eq:zdecoupl}}
z_{\ast} &=& 1048 \left[ 1 + 0.00124 (\Omega_{b} h^{2})^{-0.738}\right] \times   \nonumber \\
&& \left(1+g_{1} (\Omega_{m} h^{2})^{g_{2}} \right)  \,,
\end{eqnarray}
where the factors $g_1$ and $g_2$ are given by
\begin{eqnarray}
g_{1} &=& \frac{0.0783 (\Omega_{b} h^{2})^{-0.238}}{1+39.5(\Omega_{b} h^{2})^{-0.763}}\,, \nonumber \\
g_{2} &=& \frac{0.560}{1+21.1(\Omega_{b} h^{2})^{1.81}} \,.
\end{eqnarray}

Finally, $r(z_{\ast}, \boldsymbol{p})$ is the comoving distance at decoupling, i.e. using the definition of the comoving distance:
\begin{equation}\label{eq:comovdist}
D_{M}(z,\boldsymbol{p})=\int_{0}^{z} \frac{c\, dz'}{H(z',\boldsymbol{p})} \; ,
\end{equation}
we set $r(z_{\ast},\boldsymbol{p}) = D_M(z_{\ast},\boldsymbol{p})$.

\subsection{Baryon Acoustic Oscillations} 
The distribution of galaxies displays detectable features which are amenable to their use as as standard rulers. Their origin are baryon Acoustic Oscillations (BAO), which produce a  pattern of
oscillations related to the physics of fluctuations in the density of visible baryonic matter produced by acoustic
density waves in the primordial plasma. Specifically the signal is encoded in the 
maximum distance that acoustic waves can travel before
the plasma gets cooled at the recombination moment with the simultaneous freezing of the wave.

The most updated data set of Baryon Acoustic Oscillations (BAO) is made of results from different surveys. In general, the $\chi_{\rm BAO}^2$ is defined as
\begin{equation}
\chi^2_{\rm BAO} = \Delta \boldsymbol{\mathcal{F}}^{\rm BAO} \, \cdot \ \mathbf{C}^{-1}_{\rm BAO} \, \cdot \, \Delta  \boldsymbol{\mathcal{F}}^{\rm BAO} \ ,
\end{equation}
with the observables $\mathcal{F}^{\rm BAO}$ changing from survey to survey.

The WiggleZ Dark Energy Survey \cite{Blake:2012pj} provides, at redshifts $z=\{0.44, 0.6, 0.73\}$, the acoustic parameter
\begin{equation}\label{eq:AWiggle}
A(z,\boldsymbol{p}) = 100  \sqrt{\Omega_{m} \, h^2} \frac{D_{V}(z,\boldsymbol{p})}{c \, z} \, ,
\end{equation}
with $h=H_0/100$, and the Alcock-Paczynski distortion parameter
\begin{equation}\label{eq:FWiggle}
F(z,\boldsymbol{p}) = (1+z)  \frac{D_{A}(z,\boldsymbol{p})\, H(z,\boldsymbol{p})}{c} \, ,
\end{equation}
where $D_{A}$ is the angular diameter distance defined as
\begin{equation} \label{eq:ang_dist}
D_{A}(z,\boldsymbol{p})=\frac{1}{1+z}\int_{0}^{z} \frac{c\, dz'}{H(z',\boldsymbol{p})} \;,
\end{equation}
and
\begin{equation}
D_{V}(z,\boldsymbol{\theta})=\left[ (1+z)^2 D^{2}_{A}(z,\boldsymbol{\theta}) \frac{c z}{H(z,\boldsymbol{\theta})}\right]^{1/3}
\end{equation}
is the geometric mean of the radial and tangential BAO modes.

The latest release of the Sloan Digital Sky Survey (SDSS) Extended Baryon Oscillation Spectroscopic Survey (eBOSS) observations \cite{Tamone:2020qrl,deMattia:2020fkb,BOSS:2016wmc,Gil-Marin:2020bct,Bautista:2020ahg,Nadathur:2020vld,duMasdesBourboux:2020pck,Hou:2020rse,Neveux:2020voa} provides:
\begin{equation}
\frac{D_{M}(z,\boldsymbol{p})}{r_{s}(z_{d},\boldsymbol{p})}, \qquad \frac{c}{H(z,\boldsymbol{p}) r_{s}(z_{d},\boldsymbol{p})} \,,
\end{equation}
where the comoving distance $D_M$ is given by Eq.~(\ref{eq:comovdist}) and the sound horizon is evaluated at the dragging redshift $z_{d}$. The dragging redshift is estimated using the analytical approximation provided in  \cite{Eisenstein:1997ik} which reads
\begin{equation}\label{eq:zdrag}
z_{d} = \frac{1291 (\Omega_{m} \, h^2)^{0.251}}{1+0.659(\Omega_{m} \, h^2)^{0.828}} \left[ 1+ b_{1} (\Omega_{b} \, h^2)^{b2}\right]\; ,
\end{equation}
where the factors $b_1$ and $b_2$ are given by
\begin{eqnarray}\label{eq:zdrag_b}
b_{1} &=& 0.313 (\Omega_{m} \, h^2)^{-0.419} \left[ 1+0.607 (\Omega_{m} \, h^2)^{0.6748}\right] \,,
	\nonumber \\
b_{2} &=& 0.238 (\Omega_{m} \, h^2)^{0.223}\,.
\end{eqnarray}

Data from \cite{Zhao:2018gvb} are instead expressed in terms of 
\begin{equation}
D_{A}(z,\boldsymbol{p}) \frac{r^{fid}_{s}(z_{d},)}{r_{s}(z_{d},\boldsymbol{p})}, \qquad H(z,\boldsymbol{p}) \frac{r_{s}(z_{d},\boldsymbol{p})}{r^{fid}_{s}(z_{d},\boldsymbol{p})} \,,
\end{equation}
where $r^{fid}_{s}(z_{d})$ is the sound horizon at dragging redshift calculated for the given fiducial cosmological model considered in \cite{Zhao:2018gvb}, which is equal to $147.78$ Mpc.

\subsection{Statistical tools}

The total $\chi^2$ combining contributions from each data set is minimized using our own code for Monte Carlo Markov Chains (MCMC). The convergence of the chains is checked using the diagnostic described in \citep{Dunkley:2004sv} . 

We compare our model to two standard scenarios,  $\Lambda$CDM and  quiessence ($w={\rm const}).$ For a well-based statistical comparison, we calculate the Bayes Factor \cite{bayesfactors}, $\mathcal{B}^{i}_{j}$, defined as the ratio between the Bayesian Evidences of model $\mathcal{M}_i$ (in our case the quiessence and the singularity model) and the model $\mathcal{M}_j$ assumed as reference, in this case being the $\Lambda$CDM. We calculate the evidence numerically using our own code implementing the Nested Sampling algorithm developed by \cite{Mukherjee:2005wg}. Finally, the interpretation of the Bayes Factor is conducted using the empirical Jeffrey's scale \cite{Jeffreys1939-JEFTOP-5}: $\ln \mathcal{B}^{i}_{j}<1$ means inconclusive (strength of) evidence;  $1<\ln \mathcal{B}^{i}_{j}<2.5$ indicates weak evidence; $2.5<\ln \mathcal{B}^{i}_{j}<5$ points to moderate evidence; and $\ln \mathcal{B}^{i}_{j}>5$ is strong evidence.


{\renewcommand{\tabcolsep}{2.5mm}
{\renewcommand{\arraystretch}{2.}
\begin{table*}
\begin{minipage}{\textwidth}
\huge
\centering
\caption{Results from the statistical analysis. For each parameter we provide the median and the $1\sigma$ constraints; fixed or derived parameters are in italic font; unconstrained parameters are in typewriter font (in this case we provide both median and value at the minimum $\chi^2$ within brackets). The columns show: $1.$ considered theoretical scenario; $2.$ dimensionless matter parameter, $\Omega_m$; $3.$ dimensionless baryonic parameter, $\Omega_b$; $4.$ dimensionless Hubble constant, $h$; $5.$ absolute magnitude calibration point for SNeIa, $\mathcal{M}$; $6.$ dimensionless dark energy parameter (secondary parameter, for $\Lambda$CDM and quiessence models; primary parameter for all the other cases); $7.$ dark energy equation of state parameter $w$ for $\Lambda$CDM and quiessence models, $\eta$ parameter for all other cases; $8.$ minimum of the total $\chi^2$; $9.$ asymptotic acceptance ratio of the MCMCs; $10.$ logarithm of the Bayes factor $\log \mathcal{B}_{i}^{j}$.}\label{tab:results_1}
\resizebox*{0.95\textwidth}{!}{
\begin{tabular}{c|cccc|cc|ccc}
\hline
\hline
 & $\Omega_m$ & $\Omega_b$ & $h$ & $\mathcal{M}$ & $\Omega_{\Lambda}$ 
& $w,\eta$ & $\chi^{2}_{\rm min}$ & accept. & $\log \mathcal{B}^{j}_{i}$\\
\hline
\hline
$\Lambda$CDM & $0.318^{+0.006}_{-0.006}$ & $0.0493^{+0.0006}_{-0.0006}$ & $0.674^{+0.004}_{-0.004}$ & $-19.44^{+0.01}_{-0.01}$ & $\mathit{0.682^{+0.006}_{-0.006}}$ & $\mathit{-1.}$ & $1648.33$ & $0.31$ & $\mathit{0.}$ \\
quiessence & $0.315^{+0.006}_{-0.006}$ & $0.0478^{+0.0009}_{-0.0008}$ & $0.683^{+0.006}_{-0.006}$ & $-19.42^{+0.01}_{-0.01}$ & $\mathit{0.685^{+0.006}_{-0.006}}$ &  $-1.071^{+0.030}_{-0.031}$ & $1642.61$ & $0.26$ & $2.49^{+0.03}_{-0.04}$ \\
\hline
\hline
$\Omega_{\Lambda}>0$ & $0.315^{+0.006}_{-0.006}$ & $0.0478^{+0.0009}_{-0.0009}$ & $0.683^{+0.005}_{-0.006}$ & $-19.42^{+0.01}_{-0.01}$ & $\mathtt{<7.78\, (0.721)}$ & $0.228^{+0.105}_{-0.095}$ & $1641.46$ & $0.07$ & $2.54^{+0.03}_{-0.03}$ \\
$\Omega_{\Lambda}=0.7$ & $0.316^{+0.005}_{-0.006}$ & $0.0473^{+0.0009}_{-0.0008}$ & $0.686^{+0.006}_{-0.006}$ & $-19.41^{+0.01}_{-0.01}$ & $\mathit{0.7}$ & $1.268^{+0.313}_{-0.319}$ & $1638.82$ & $0.15$ & $4.38^{+0.03}_{-0.02}$ \\
$\Omega_{\Lambda}=1$ & $0.315^{+0.006}_{-0.006}$ & $0.0478^{+0.0009}_{-0.0008}$ & $0.684^{+0.006}_{-0.006}$ & $-19.42^{+0.01}_{-0.01}$ & $\mathit{1.}$ & $0.349^{+0.138}_{-0.141}$ & $1642.27$ & $0.29$ & $2.74^{+0.03}_{-0.03}$\\
$\Omega_{\Lambda}=5$ & $0.315^{+0.006}_{-0.006}$ & $0.0478^{+0.0009}_{-0.0008}$ & $0.683^{+0.006}_{-0.006}$ & $-19.42^{+0.01}_{-0.01}$ & $\mathit{5.}$ & $0.218^{+0.091}_{-0.088}$ & $1642.58$ & $0.28$ & $2.51^{+0.03}_{-0.03}$\\
\hline
\hline
$\Omega_{\Lambda}>0,\, \eta =2.5$ & $0.320^{+0.006}_{-0.006}$ & $0.0469^{+0.0008}_{-0.0008}$ & $0.688^{+0.005}_{-0.005}$ & $-19.40^{+0.01}_{-0.01}$ & $0.681^{+0.006}_{-0.006}$ & $\textit{2.5}$ & $1636.68$ & $0.18$ & $5.22^{+0.03}_{-0.03}$\\
$\Omega_{\Lambda}>0,\, \eta =1$ & $0.316^{+0.006}_{-0.006}$ & $0.0474^{+0.0008}_{-0.0008}$ & $0.686^{+0.005}_{-0.005}$ & $-19.41^{+0.01}_{-0.01}$ & $0.719^{+0.027}_{-0.019}$ & $\textit{1.}$ & $1639.77$ & $0.25$ & $3.92^{+0.04}_{-0.03}$\\
$\Omega_{\Lambda}>0,\, \eta =0.1$ & $0.316^{+0.006}_{-0.006}$ & $0.0486^{+0.0006}_{-0.0006}$ & $0.678^{+0.004}_{-0.004}$ & $-19.43^{+0.01}_{-0.01}$ & $\mathtt{<36.54\, (47.86)}$ & $\textit{0.1}$ & $1644.17$ & $0.28$ & $2.06^{+0.03}_{-0.03}$\\
$\Omega_{\Lambda}>0,\, \eta =0.001$ & $0.318^{+0.006}_{-0.006}$ & $0.0493^{+0.0006}_{-0.0006}$ & $0.674^{+0.004}_{-0.004}$ & $-19.44^{+0.01}_{-0.01}$ & $\mathtt{<51.93\, (5.41)}$ & $\textit{0.001}$ & $1648.27$ & $0.15$  & $-0.006^{+0.037}_{-0.027}$\\
$\Omega_{\Lambda}>0,\, \eta =-0.5$ & $0.319+^{+0.006}_{-0.006}$ & $0.0497^{+0.0007}_{-0.0007}$ & $0.672^{+0.004}_{-0.004}$ & $-19.44^{+0.012}_{-0.012}$ & $0.700^{+0.026}_{-0.015}$ & $\textit{-0.5}$ & $1648.53$ & $0.32$  & $-0.88^{+0.03}_{-0.03}$\\
$\Omega_{\Lambda}>0,\, \eta =-1$ & $0.319+^{+0.006}_{-0.006}$ & $0.0497^{+0.0007}_{-0.0007}$ & $0.6720^{+0.004}_{-0.004}$ & $-19.44^{+0.012}_{-0.012}$ & $0.693^{+0.015}_{-0.010}$ & $\textit{-1.}$ & $1648.43$ & $0.29$ & $-0.76^{+0.03}_{-0.03}$ \\
\hline
\hline
\end{tabular}}
\end{minipage}
\end{table*}}}

\section{Results and Discussion}
\label{sec:results}

Results from our statistical analysis are reported in Table~\ref{tab:results_1}, where full $1\sigma$ constraints on all parameters involved are reported, together with the minimum value of the $\chi^2$ and the Bayes Factors.

We first note that performing an analysis wheere both  model parameters $\{\Omega_{\Lambda}, \eta\}$ are left free has revealed problematic. Due to the high correlations among the two, and because of the ``asymptotic'' nature of the mathematical function used to describe the future singularity, the constraints are poor (statistically speaking) if both  parameters are allowed to vary simultaneously. Thus, in order to get more information about the $\chi^2$ landscape, we have decided to consider as many cases as possible in which one parameter is fixed (assuming various different values) and the other one is left free to vary. In this way, we can get information on the more general behaviour of the $\chi^2$ in the full parameter space.

We can see that when  $\Omega_{\Lambda}$ varies,  a clear indication emerges for a preference over values $\sim 0.7$. Departing from it, both the $\chi^2$ and the Bayes Factor degrade quite fast. To this $\Omega_{\Lambda} \sim 0.7$ value we can  associate a quite well constrained estimation of the other free parameter, $\eta\sim 1.268$. This general trend seems to be confirmed also when we  fix $\eta$ instead and leave $\Omega_{\Lambda}$ free. In this case we see that the best $\chi^2$ and Bayes Factor correspond to $\eta \sim 2.5$ and $\Omega_{\Lambda} \sim 0.681$. We can also note how large relative variations of $\eta$ (for $\eta>0$) lead to small variations in $\Omega_{\Lambda}$, which is one of the reasons which make the fully-free analysis difficult. On the other hand, values of $\eta<1$ seem to be strongly discarded from a statistical point of view, with very small values leading to even negative values of the Bayes Factor, which means they are totally disfavoured.

If we focus on the scenario which turns out to present a moderate evidence in favor of the ripped $\Lambda$CDM model, i.e. the one with $\eta=2.5$, it would be interesting to check in detail which probe is responsible for the significant improvement in the statistics. Indeed, the $\chi^2$ is lowered by $\approx12$ units with respect to the reference $\Lambda$CDM case. Dissecting the total $\chi^2$ by probe, we see how the largest improvement comes from the SNeIa, whose $\chi^{2}_{SN}$ improves by $\approx11$ units, followed by BAO SDSS-IV DR16 Lyman $\alpha$ data \cite{duMasdesBourboux:2020pck} and Quasars both from the latest DR16 release \cite{Hou:2020rse,Neveux:2020voa} and from the DR14B one \cite{Zhao:2018gvb}, with respectively produce an improvement of $\approx3$, $\approx1.8$ and $\approx 1.8$ units. All other probes do not have statistically significant differences, except for the CMB shift parameters, whose $\chi^2$ worsen by $\approx4$ units. It is interesting to note that BAO data also depend on early-times quantities, but while CMB data  generally fit our model worse, in the case BAO improvement occurs. Thus, we may safely conclude that the ripped $\Lambda$CDM greatly and mostly improves in a more decisive way the description of late time $(z \leq 2.5)$ data than that at higher redshifts. 

\begin{figure*}[htbp!]
\centering
\includegraphics[width=8.5cm]{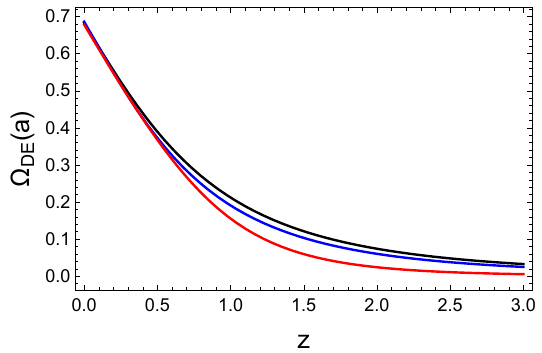}~~~
\includegraphics[width=8.5cm]{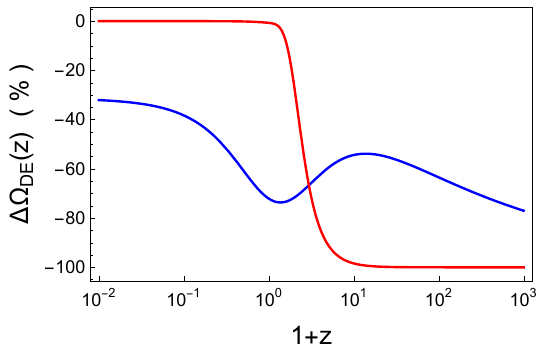}\\
~~~\\
\includegraphics[width=8.5cm]{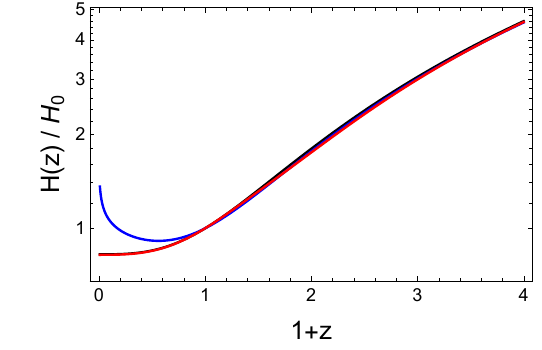}~~~
\includegraphics[width=8.5cm]{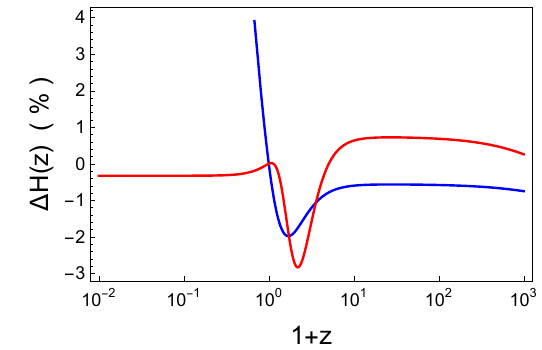}
\caption{Comparison of the time behaviour of dark energy component $\Omega_{DE}(a)$ (top panels), and of the dimensionless Hubble parameter $H(z)/H_0$ (bottom panels) for three different scenarios: black - $\Lambda$CDM; blue - $w=const.$; red - ripped $\Lambda$CDM with $\eta = 2.5$. The curves are drawn using the parameter values minimizing the $\chi^2$; errors intervals are omitted just for the sake of clarity. On the right columns, we plot differences $100(X_{model}-X_{\Lambda CDM}/X_{\Lambda CDM}$).}
\label{fig: Omega_DE}
\end{figure*}

This can be also clearly seen in Fig.~\ref{fig: Omega_DE} where we show the behaviour of the dimensionless dark energy component for three cases which we have considered in our analysis. The largest departure of our ripped $\Lambda$CDM model from a standard $\Lambda$CDM one is exactly in the range covered by the probes described above, while in the future we asymptotically approach $\Lambda$CDM, and in the early times we need much less dark energy than what required for a cosmological constant. 

Moreover, looking at the behaviour of the Hubble parameter, we have another confirmation that the deviations of our model from the standard $\Lambda$CDM are minimal, being less than $3\%$ both at early and late times, and even in future epochs.

It is also interesting to note that, looking at Eq.~(\ref{eq:density_lambda}), we can see how the combination $\lambda^{-1/\eta}$ plays the role of an \textit{effective} scale factor where the transition which characterizes our model is turned on. From our MCMC results, we infer for such transition scale factor a median value of $0.46^{+0.02}_{-0.03}$ or equivalently, a transition redshift $1.16^{+0.12}_{-0.10}$, which falls in the range where our model seems to perform better than a standard $\Lambda$CDM.

Finally, in Fig.~\ref{fig:w_eff}, we show the evolution of the equation of state for our model, from $z=0$ to $z=3$ in $0.25-$steps. We clearly have a phantom behaviour all over the range, approaching asymptotically a cosmological constant only at recent times.

\begin{figure}[htbp!]
\begin{center} 
\animategraphics[controls,width=8cm]{1}{gif/prova_}{1}{13}
\caption{Dynamical visualization of the equation of state parameter, $w_(z)$ sampled from $z=0.$ to $z=3$, with fixed $0.25$ steps.}\label{fig:w_eff}
\end{center} 
\end{figure}


\section{Conclusions}
\label{sec:conclucions}

Predicting the fate of the Universe may seem philosophy, but it is quite a physical strive. A battery of background astrophysical and cosmological probes may assist us in giving a strong picture of the dynamics of the Universe from close to its very beginning to the current moment. An established practice in the literature assists this endeavour by exploring dark energy parametrizations with a small set of otherwise powerful assumptions.

In particular tight restrictions can be on the degrees of freedom of a representative of a new family of transient evolutions defined through a specific functional form on its dark energy density. A detailed discussion confirms that our study case (based on a sigmoid function), Eq.~(\ref{eq:density_lambda}) reproduces an intriguing new cosmological model with phantom behaviour all throughout its history to evolve very smoothly towards an asymptotic non-singular end in the form of a pseudorip; that is, the energy density evolves into the $p=-\rho$ condition in infinite time. The characteristics of the model suggests coining a new denomination for it: Ripped $\Lambda$CDM. 

For our observation-based tests we choose a sufficiently reliable combination of background data sets with well-known cosmic complementarity, which is in turn a route leading to the reduction of uncertainty on cosmological fits.


At first level these tools, through a thorough statistical analysis, let us join the consensus on the evidence of a large scale accelerated expansion driven by a mysterious power. As could not be otherwise, our model offers yet another piece of evidence of the everyday increasing rate of expansion in the universe.

But at the next level our examinations reveal some surprising substantial preference (established in terms of rigorous statistical tools like the Evidence and the Bayes Factor) for models with  a pseudorip finale over $\Lambda$CDM (the so-called concordance model). As this result comes from the use of a specific choice of a smooth transition between a phantom-like dark energy to a cosmological constant one, it would be very interesting to continue to reproduce the whole analysis with similar model with either a faster or a slower transition.


Our tests pinpoint the order of magnitude  of the observationally preferred value of one the free parameters of the model: $\eta \sim {\cal O}(1)$. By  fixing this parameter our model would become an interesting one-parameter (free $\Omega_{\lambda}$) evolutionary parametrization of the dark energy to be tested in further depth. Note that choosing the value of one parameter would lessen statistical uncertainty to some extent and would allow us to be more certain about the preminence of new parametrizations of this form.

Remarkably, if we chose to reinterpret our models in terms of a phantom scalar field and examine its late time bounds as given by the best evidence offering values of $\eta$ among those considered we find that there is very good compatibility ($\phi=1.79$ for $\eta=2.5$) with swampland distance conjecture restrictions, and this can be taken as extra motivation to continue to pursue these and similar models.

New insights derived from additional tests would certainly contribute to strengthening our knowledge about how seriously we can take the possibility of peaceful yet awkward destiny of our universe. 

\section*{Acknowledgements }
RL and MBL are supported by the Basque Government Grant IT1628-22.  RL is also supported by Grant PID2021-123226NB-I00 (funded by MCIN/AEI/10.13039/501100011033 and by “ERDF A way of making Europe”). This article is based upon work from COST Action CA21136 Addressing observational tensions in cosmology with systematics and fundamental physics (CosmoVerse) supported by COST (European Cooperation in Science and Technology). MBL is also supported by the Basque Foundation of Science Ikerbasque and by Grant PID2020-114035GB-100 (MINECO/AEI/FEDER, UE). RL is also supported by Generalitat Valenciana through research project PROMETEO/2020/079.
\bibliographystyle{apsrev4-1}
%

\end{document}